\documentclass[12pt]{article}
\usepackage{latexsym,amssymb,amsmath}

\begin{document}

\title
{Multi-instantons and superstring solitons}
\author
{E.K. Loginov\footnote{{\it E-mail address:} loginov@ivanovo.ac.ru}
\medskip\\
\it Department of Physics, Ivanovo State University\\
\it Ermaka St. 39, Ivanovo, 153025, Russia}
\date{}
\maketitle

\begin{abstract}
Multi-instanton solutions in the eight and seven dimensional Yang-Mills fields theory is obtained. Extended-soliton solutions to the low-energy heterotic-field-theory equations of motion is constructed from this higher-dimensional multi-instantons.
\end{abstract}

\bigskip

In the past few years, classical solitonic solutions in string theory with higher-brane structure have been actively investigated. These solutions are static multi-soliton solutions obeying a zero-force condition and saturating a Bogomol'nyi bound between ADM mass and charge. In certain cases~(See Ref.~[1]), exact fivebrane solutions of heterotic string theory may be constructed, each solution in principle corresponding to an exact conformal field theory of the sigma-model. Although the solutions are initially conceived as perturbative expansions in the classical string parameter $\alpha'$, the exact solutions acquire nonperturbative status. Being classical, the solitons are tree-level solutions in the quantum topological expansion of the string worldsheet, but are also nonperturbative in the loop parameter. Therefore, full quantum string-loop extensions of these solutions await an understanding of nonperturbative string theory. Nevertheless, it is possible to use these solitons in nonperturbative calculations, such that vacuum-tunneling, since it is often the case that higher order corrections do not contribute to these effects.
\par 
In Ref.~[2] a one-brane solution of heterotic theory was found which is an everywhere smooth solution of the equations of motion. The construction of this solution involves crucially the properties of octonions. One of the many bizarre features of this soliton is that it preserves only one of the sixteen space-time supersymmetries, in contrast to previously known examples of supersymmetric solitons which all preserve half of the supersymmetries. In Ref.~[3] a two-brane solution of heterotic theory was found. This soliton preserves two of the sixteen supersymmetries and hence corresponds to $N=1$ space-time supersymmetry in $(2+1)$ dimensions transverse to the seven dimensions where the Yang-Mills instanton is defined. 
\par
As in these Refs. we search for a solution to lowest nontrivial order in $\alpha'$ of the equations of motion that follow from the bosonic action
\begin{equation}
S=\frac{1}{2k^2}\int d^{10}x\,\sqrt{-g}e^{-2\phi}\left(R+4(\nabla\phi)^2
-\frac{1}{3}H^2-\frac{\alpha'}{30}\text{Tr}F^2\right),
\end{equation}
where the three-form antisymmetric field strength is related to the two-form potential by the familiar anomaly equation
\begin{equation}
H=dB+\alpha'\left(\omega_3^{L}(\Omega)-\frac{1}{30}\omega_3^{YM}(A)\right)+\dots,
\end{equation}
where $\omega_3$ is the Chern-Simons three-form and the connection $\Omega_{M}$ is a non-Riemannian  connection related to the usual spin connection $\omega$ by
\begin{equation}
\Omega_{M}^{AB}=\omega_{M}^{AB}-H_{M}^{AB}.
\end{equation}
We are interested in solutions that preserve at least one supersymmetry. This requires that in ten dimensions there exist at least one Majorana-Weyl spinor $\epsilon$ such that the supersymmetry variations of the fermionic fields vanish for such solutions
\begin{align}
\delta\chi&=F_{MN}\Gamma^{MN}\epsilon\nonumber,\\  
\delta\lambda&=(\Gamma^{M}\partial_{M}\phi-\frac16H_{MNP}\Gamma^{MNP})\epsilon,\\
\delta\psi_{M}&=(\partial_{M}+\frac14\Omega_{M}^{AB}\Gamma_{AB})\epsilon\nonumber.
\end{align}
\par\smallskip
We start by picking definite commuting $SO(8)$ spinor $\eta$ with $\Gamma_9\eta=\eta$ normalized to $\eta^{T}\eta=1$. We can then introduce a fourth-rank antisymmetric $Spin(7)$-invariant tensor
\begin{equation}
f_{mnps}=-\eta^{T}\Gamma_{mnps}\eta.
\end{equation}
There exists an explicit construction of the $SO(8)$ gamma matrices in terms of the octonion structure constants $c_{ijk}$ defined by
\begin{equation}
e_{i}e_{j}=-\delta_{ij}+c_{ijk}e_{k},
\end{equation}
where $c_{ijk}$ are antisymmetric in $(i,j,k)$ and nonzero and equal to unity for the seven combinations $(123)$, $(145)$, $(167)$, $(246)$, $(275)$, $(374)$, $(365)$. Using this construction and an explicit choice for $\eta$ one finds
\begin{align}
f_{ijk8}&=c_{ijk},\nonumber\\
f_{ijkl}&=\delta_{il}\delta_{jk}-\delta_{ik}\delta_{jl}+c_{ijr}c_{klr}.
\end{align}
Suppose $G_{\alpha}=SU(2)$, $G_2$, $Spin(7)$ and $d_{\alpha}=4,7,8$ for $\alpha=1,2,3$, respectively, and define the projectors $f^{\alpha}_{mnps}$ of the Lie algebra $so(8)$ onto the subalgebras $so(3)$, $g_2$ and $so(7)$ by
\begin{eqnarray}
f^{\alpha}_{mnps}=\frac{1}{2(\alpha+1)}[\alpha(\delta_{mp}\delta_{ns}-\delta_{ms}\delta_{np})-f_{mnps}].
\end{eqnarray}
Noting that the indices $m,n,p,s$ in (8) take values from $1$ to $d_{\alpha}$. Suppose
\begin{eqnarray}
E^{\alpha}_{mn}=f^{\alpha}_{mnps}E_{ps},
\end{eqnarray}
where $E_{mn}$ are the generators of $so(8)$ satisfying the switching relations
\begin{eqnarray}
[E_{mn},E_{ps}]=\delta_{p[m}E_{n]s}-\delta_{s[m}E_{n]p}.
\end{eqnarray}
Using the well-known identities for $f_{mnps}$ (see, e.g., Ref.~[4]), we get the switching relations
\begin{multline}
[E^{\alpha}_{mn},E^{\alpha}_{ps}]=\frac{1}{2(\alpha+1)}[2\alpha(E^{\alpha}_{m[p}\delta_{s]n}-
E^{\alpha}_{n[p}\delta_{s]m})\\
-f_{mnk[p}E^{\alpha}_{s]k}+f_{psk[m}E^{\alpha}_{n]k}].
\end{multline}
\par 
Consider now the Yang-Mills gauge theory in $d_{\alpha}$ dimensions with the gauge group $G_{\alpha}$. We proceed from the following ansatz
\begin{eqnarray}
A_m(x)=\frac{\alpha+1}{\alpha}\frac{\lambda^{\dag}y^{n}}{1+y^{\dag}y}E^{\alpha}_{mn},
\end{eqnarray}
where $y$ is a column vector with the octonions $y_1,\dots,y_{N}$ such that
$$
\begin{aligned}
y^{\dag}&=(y^{k}_1,\dots,y^{k}_{N})\bar e_{k},&\qquad y^{k}_{I}&\in\mathbb R,\\
\lambda^{\dag}&=(\lambda_1,\dots,\lambda_{N}),&\qquad \lambda_{I}&\in\mathbb  R^+,\\
y^{k}_{I}&=(b^{k}_{I}+x^{k})\lambda_{I},
\end{aligned}
$$
where we do not sum on the recurring indices $I$. Using the switching relations (11), we get the field strength
\begin{eqnarray}
F_{mn}=-\frac{\alpha+1}{\alpha^2}\frac{\lambda^{\dag}\{\alpha(2+2y^{\dag}y-y^{i}y_{i}^{\dag})
E^{\alpha}_{mn} +2(\alpha+1)f^{\alpha}_{mnis}E^{\alpha}_{sj}y^{j}y_{i}^{\dag}\}\lambda}{(1+y^{\dag}y)^2}.
\end{eqnarray}
It is easy to verify that the solutions (12) satisfy the Yang-Mills field equations by virtue of the Bianchi identities and the self-duality of the field strength (13).
\par
If $\alpha=1$ then the Euclidean Yang-Mills action is finite. In the case the topological charge is equal to the unity and the obtained solution is gauge equivalent to the instanton solution of Ref.~[5]. If $\alpha\ne1$ then the Yang-Mills action is infinite and the topological meaning of the solutions in unknown. In the case we cannot assert that the solutions are gauge equivalent to the instanton solutions that was found in Ref.~[6]. They are new solutions which depend on at most $(d_{\alpha}+1)N$ effective parameters. The fact that these solutions do not have finite action involve that the corresponding supersymmetric solutions do not have finite energy per unit length and it further complicates its physical implications. This is a characteristic feature of such solutions, though. There are no finite-action solutions of the Yang-Mills equations in eight or seven dimensions~(See Ref.~[7]). In order to get finite-action Yang-Mills equations, one would have to "compactify" some of the dimensions. 
\par\smallskip
Let us now show that the above multi-instanton solutions can be extended to a solitonic solution of the heterotic string. Consider the action of the ten dimensional low energy effective theory of the heterotic string. The bosonic part of this action is (1). Denote world indices of the $d_{\alpha}$-dimensional space transverse to the $(9-d_{\alpha})$-brane by $\mu,\nu=1\dots d_{\alpha}$ and the corresponding tangent space indices by $m,n=1\dots d_{\alpha}$. We assume that no fields depend on the longitudinal coordinates and that the nontrivial tensor fields in the solution have only transverse indices. Then the gamma matrix terms in (4) are sensitive only to the $G_{\alpha}$ part of $\epsilon$. Thus taking $\epsilon$ to be a $G_{\alpha}$ singlet $\eta$ and the non-vanishing components of $F_{MN}$ to be those given by the $d_{\alpha}$ dimensional octonionic multi-instanton the supersymmetry variation $\delta\chi$ of the gaugino  vanishes. This follows from the fact that
\begin{eqnarray}
f^{\alpha}_{mnps}\Gamma^{ps}\eta=0
\end{eqnarray}
and the self-duality of $F_{\mu\nu}$. To deal with the other supersymmetry variations, we must adopt an ansatz for the non-trivial behavior of the metric and antisymmetric tensor fields in the $d_{\alpha}$ dimensions transverse to the $(9-d_{\alpha})$-brane. For the antisymmetric tensor field strength we write
\begin{eqnarray}
H_{\mu\nu\lambda}=\beta f_{\mu\nu\lambda\sigma}\partial^{\sigma}\phi,
\qquad \beta=\frac{1}{3\alpha-2},
\end{eqnarray}
where $\phi$ is to be identified with the dilaton field. Using standard $d_{\alpha}$-dimen\-sional gamma-matrix algebra, we prove that the dilatino variation $\delta\lambda $ vanishes. For the metric tensor we write
\begin{eqnarray}
g_{\mu\nu}=e^{{2\alpha\beta}\phi}\delta_{\mu\nu}.
\end{eqnarray}
With this ansatz and the rather obvious vierbein choice $e^{m}_{\mu}=\delta^{m}_{\mu}e^{\alpha\beta\phi}$, we can also calculate the generalized spin connection (3) which appear in (4)
\begin{eqnarray}
\Omega_{\mu mn}=\beta[\alpha(\delta_{\mu m}\delta_{\nu n}-\delta_{\mu n}\delta_{\nu m})
-f_{\mu mn\nu}]\partial^{\nu}\phi.
\end{eqnarray}
It follows from (8) and (14) that it suffices to take $\epsilon$ to be a constant $G_{\alpha}$ invariant spinor to make the gravitino variation $\delta\psi_{M}$ vanish. Putting all this together, we see that if we choose the gauge field to be any instanton and fix the metric and antisymmetric tensor in terms of the dilaton according to the above ansatz, then the state is annihilated by all supersymmetry variations generated by a space-time constant $G_{\alpha}$ invariant spinor.
\par 
In order to fully determine the solution it remain to solve the Bianchi identity
\begin{eqnarray}
dH=\alpha'\left(\text{tr}R\wedge R-\frac{1}{30}\text{Tr}F\wedge F\right),
\end{eqnarray}
where $\text{Tr}$ refers to the trace in the adjoint representation of $E_8$ or $SO(32)$ in the corresponding heterotic string theory. This equation can be solved perturbatively in $\alpha'$ beginning with the instanton solution for. $F$. Equation (18) then implies that $R\sim O(\alpha')$. Therefore, to leading order in $\alpha'$, one is entitled to drop the $R\wedge R$ term in (18). Substituting the gauge field strength (13), one obtains the following dilaton solution:
\begin{eqnarray}
e^{-\alpha\beta\phi}=e^{-\alpha\beta\phi_0}+\frac{8(\alpha+1)}{3\alpha}
\frac{\lambda^{\dag}(2+2y^{\dag}y-y^{i}y_{i}^{\dag})\lambda}{(1+y^{\dag}y)^2}+O(\alpha'{}^2).
\end{eqnarray}
The metric and antisymmetric field are built out of this dilaton field according to the spacetime-supersymmetric ansatz of (15) and (16). However, just like the solutions of Ref.~[2] and [3], the metric and the field strength fall of only as $1/r^2$ and this implies that the ADM mass per unit length of this string diverges. Nevertheless, the correct physical interpretation of these solutions and its implications for superstring theory remain to be understood fully.  
\par\smallskip
{\it Acknowledgement:} Research supported by RFBR Grant 04-02-16324.


\small
\begin{thebibliography}{12}
\bibitem{1}
A.~Strominger, Nucl. Phys. B343 (1990) 167;\\
M.J.~Duff and J.X.~Lu, Nucl. Phys. B354 (1991) 141;\\
C.~Callan, J.~Harvey and A.~Strominger, Nucl. Phys. B359 (1991) 611.
\bibitem{2}
J.A.~Harvey, A.~Strominger, Phys. Rev. Lett. 66 (1991) 549.
\bibitem{3}
M.~Gunaydin, H.~Nicolai, Phys. Lett. B351 (1995) 169.
\bibitem{4}
R.~Dundarer, F.~Gursey and C.-H.~Tze, J. Math. Phys. 25 (1984) 1496.
\bibitem{5}
A.A.~Belavin, A.M.~Polyakov, A.S.~Schwartz and Yu.S.~Tyupkin, Phys. Lett. B59 (1975) 85.
\bibitem{6}
D.B.~Fairlie, J.~Nuyts, J. Phys. A17 (1984) 2867;\\
S.~Fubini, H.~Nicolai, Phys. Lett. B155 (1985) 369;\\
T.A.~Ivanova, A.D.~Popov, Lett. Math. Phys. 24 (1992) 85;\\
E.K.~Loginov, J. Phys. A37 (2004) 6599.
\bibitem{7}
A.~Laffe and C.~Taubes, Vortice and monopoles (Birkhauser, Boston, 1980). 

\end{thebibliography}
\end{document}